\documentclass[10pt,reqno]{amsart}
\usepackage{amssymb,mathrsfs,color}
\usepackage{pinlabel}

\usepackage{amsmath, amsfonts, amsthm, verbatim, amssymb}
\usepackage{epstopdf, mathtools}
\usepackage{color}
\usepackage{esint}
\usepackage{stmaryrd}
\usepackage{graphicx}
\usepackage{xcolor} 
\usepackage{tensor}
\usepackage{slashed}
\usepackage{cite}
\mathtoolsset{showonlyrefs}

\newcommand{\bs}[1]{\boldsymbol{#1}}

\newcommand{\cl}[1]{\mathcal{#1}}

\newcommand{\al}{\alpha}

\newcommand{\eps}{\epsilon}

\newcommand{\la}{\lambda}

\newcommand{\p}{\partial}

\usepackage[multiple]{footmisc}
\numberwithin{equation}{section}

\theoremstyle{remark}

\newcommand{\tr}{\mathrm{tr}\,}

\newcommand{\rar}{\rightarrow}

\newcommand{\bbD}{\mathbb D}
\newcommand{\bbE}{\mathbb E}

\newcommand{\bbR}{\mathbb R}

\newcommand{\tens}{\otimes}
\newcommand{\scrW}{\mathscr W}

\newcommand{\msB}{\bs{\mathsf{B}}}

\newcommand{\hbF}{\hat{\bs F}}

\newcommand{\hbP}{\hat{\bs P}}

\newcommand{\hbchi}{\hat{\bs \chi}}
\newcommand{\hbv}{\hat{\bs v}}

\begin{document}

\title[Strain-gradient elastic shells]{Shell energies derived from three-dimensional isotropic strain-gradient elasticity}
\author[]{C. Balitactac, Y. Canzani, R. S. Hallyburton, \\ J. Mott, C. Rodriguez}

\begin{abstract}
We derive a class of two-dimensional shell energies for thin elastic bodies exhibiting small-length scale effects modeled via strain-gradient elasticity. Building on the final author's earlier work on plate models, the kinetic and stored surface energies arise as the leading cubic order-in-thickness expressions for three-dimensional kinetic energies with velocity-gradient effects and a broad class of isotropic stored energies, each possessing an intrinsic length scale $\ell$. These include both classical Toupin–Mindlin and more recent dilatational strain-gradient elastic stored energies. A key insight of this work is that consistent asymptotic reductions of strain-gradient theories necessarily begin at cubic order-in-thickness due to the natural scaling assumption $\ell = O(h)$ where $h$ is the thickness of the body. In the limit as the intrinsic length scales vanish, the theory reduces to Koiter’s classical shell energy. We illustrate the theory using the shell energy derived from dilatational strain-gradient elasticity, computing the body force, edge tractions and edge double force densities required to support a variety of finite deformations.
\end{abstract}

\maketitle

\section{Introduction}

\subsection{Material surfaces incorporating small-length scale properties} For decades, researchers have developed two-dimensional material surface theories to model thin three-dimensional bodies that exhibit small-length scale effects. A prominent example is the theory of \textit{fibrous networks}. In the context of fibrous networks, the geometry of the material surface is encoded by two orthonormal vector fields, $\boldsymbol L$ and $\boldsymbol M$, defined on a surface $\Omega \subset \mathbb E^3$.\footnote{In this work, we denote three-dimensional Euclidean space by $\bbE^3$ and identify its translation space with $\bbR^3$ via a fixed orthonormal basis $\{ \bs e_i \}_{i = 1}^3$. Throughout this work, we use standard vector and tensor operations and conventions in $\bbR^3$. We also raise and lower indices using the flat metric on $\bbR^3$, and we use the Einstein summation convention that repeated indices in upper and lower positions imply summation. 
	
Finally, we use standard big-oh and little-oh notation, e.g., $A = O(B)$ means that there exists $C \geq 0$ such that $|A| \leq C B$. We say that a big-oh term depends on $D$ if $C$ depends on $D$, $C = \hat C(D)$. } These fields represent tangent directions to a network of embedded fibers in the reference configuration, while their convected counterparts define the current network through their integral curves. The surface energy governing equilibrium states via Hamilton’s principle is expressed in terms of geometric and kinematic invariants of these curves including stretch, normal curvature, and torsion.

Classical treatments of such networks, notably those of Rivlin \cite{Rivlin55}, Green and Adkins \cite{GreenAdkins1970Book}, Pipkin \cite{Pipkin81}, and Steigmann and Pipkin \cite{SteigmannPipkin91}, focused on idealized fibers with no bending or torsional resistance, and surface energies depending primarily on stretch and shear. Subsequent developments introduced \textit{certain second gradient effects} to capture resistance to fiber bending and twisting. Wang and Pipkin \cite{WangPipkin86} proposed a theory for inextensible fibers with bending stiffness, later extended by Steigmann \cite{Steigmann18CrossedElastb} to incorporate fiber twist.

A more comprehensive theory was presented by Steigmann and dell’Isola \cite{SteigmanndellIsola15}, where the surface energy depends on the first and \textit{all second-order} surface derivatives of the deformation. Numerical investigations based on these theories demonstrated: regions of uniform shear flanked by thin bands of fiber bending in pantographic lattices \cite{Giorgioetal15}; geodesic buckling in shear \cite{Giorgioetal16}; bulging behavior in strain-gradient cylinders distinct from classical membrane predictions \cite{Giorgioetal18}; and edge-localized strain energies in Hypar nets \cite{Giorgioetal19}. Finally, a general equilibrium theory, including material symmetry and a virtual work principle, was developed for material surfaces with stored surface energy depending on the first and second surface derivatives of the deformation in \cite{Steigmann18Lattice}.

The previous works adopted a \textit{direct approach}, constructing surface theories without reference to parent three-dimensional models. More recently, however, Steigmann and collaborators \cite{steigmann2023cosserat, steigmann2023thin} have developed a dimensional reduction framework rooted in three-dimensional elasticity. Their focus is on elastic solids reinforced by networks of embedded fibers, a special case of nonlinear Cosserat elasticity \cite{Steigmann2012fibers, Steigmann2015effects, ShiraniSteigmann2020}. These models' governing shell energies are obtained by integrating the three-dimensional energies through the body's thickness and choosing appropriate director fields that arise naturally from a Taylor expansion of the deformation in the direction normal to the midsurface. The methodology builds on a long line of work in classical elasticity involving asymptotic expansions and reduced-order modeling \cite{Koit66, HilgPip92b, HilgPip96, HilgPip97, Steig13, SteigShir19, steigmann2023lecture}.

Another key class of materials that exhibit small-length scale effects are \textit{strain-gradient elastic solids}. The origins of this theory trace back to Piola’s work in 1846 \cite{Piola1846Book, dellIsola15}, which introduced the concept of higher-order spatial derivatives in continuum mechanics. The modern theory, however, crystallized in the mid-20th century, through foundational work by Toupin \cite{Toupin62, Toupin64}, Green and Rivlin \cite{GreenRivlin64a, GreenRivlin64b}, Mindlin \cite{Mindlin64a, Mindlin1965}, Mindlin and Eshel \cite{MindlinEshel1968}, and Germain \cite{Germain73a, Germain73b}. These efforts led to rigorous formulations of gradient continua, particularly in elasticity. For comprehensive reviews of the extensive developments and applications of this area, we refer to \cite{Askes2011, maugin2011, maugin2013, dellIsola17, Maugin17Book, dellIsola2020higher, bertram2020} and references therein.

Despite the maturity of strain-gradient elasticity, relatively little attention has been given to the development of reduced theories for thin strain-gradient elastic bodies. To our knowledge, the \textit{only existing result} is the derivation of two-dimensional \textit{plate models} from three-dimensional stored energies due to one of the present authors \cite{rodriguez2024midsurface} and focused on a specific class of isotropic strain-gradient elastic materials. In that same study, the two-dimensional model was integrated into the surface-substrate framework proposed in \cite{rodriguez2024elastic} and used to eliminate pathological singularities that arise in classical linear elastic fracture mechanics applied to mode-III far-field loading of a finite-length crack. Our discussion reveals a significant gap in the literature, one with important potential applications, and highlights the need for a systematic framework for dimensionally reduced strain-gradient models of thin elastic structures.

The present work significantly extends the results of \cite{rodriguez2024midsurface} by deriving \textit{shell energies} as the leading cubic order-in-thickness contributions to three-dimensional kinetic energies with velocity-gradient effects and a broader class of isotropic strain-gradient stored energies, each possessing intrinsic length scales. In the setting of strain-gradient stored energies, we obtain the form of the shell energy $U$ satisfying 
\begin{gather}
	\int_{\mathcal B_h} W(\tilde{\boldsymbol E}, \ell_s\mathrm{Grad}\, \tilde{\boldsymbol E}) \, dV = \int_{\Omega} U \, dA + O(h^4), \label{eq:cubic}
\end{gather}
where $W$ is the stored energy of the ``thin" body $\mathcal B_h$ with generally curved midsurface $\Omega$ and thickness $h$ (see Figure 1), $\tilde{\boldsymbol E}$ is the Lagrange strain, and $\ell_s$ is an intrinsic length scale associated to $W$.\footnote{See Section 2 for a more thorough discussion of the geometry and kinematics associated to $\mathcal B_h$ and $\Omega$.} The three-dimensional stored energies $W$ include classical Toupin–Mindlin type energies \cite{Toupin62, Toupin64, Mindlin64a, MindlinEshel1968} that incorporate the most general isotropic quadratic forms of the strain-gradients. In addition, we derive shell energies for a class of more recently introduced dilatational models \cite{eremeyev2021nonlinear}, in which the stored energy depends objectively only on the deformation gradient and the gradient of its determinant. A dilatational model can be viewed as a capillary fluid (first introduced by \cite{korteweg1901}) ``solidifying"; see Section 3 of \cite{eremeyev2021nonlinear}. Moreover, in the limiting case where the intrinsic length scale vanishes, we recover Koiter’s classical shell energy. Asymptotic expansions for kinetic energies possessing velocity gradients similar to \eqref{eq:cubic} are also obtained. Combined with Hamilton's principle, the shell energies derived in this work provide a foundation for future studies of thin structures where microstructural effects are incorporated via strain-gradient elasticity. 

\begin{figure}[t]
\includegraphics[width=.9\linewidth]{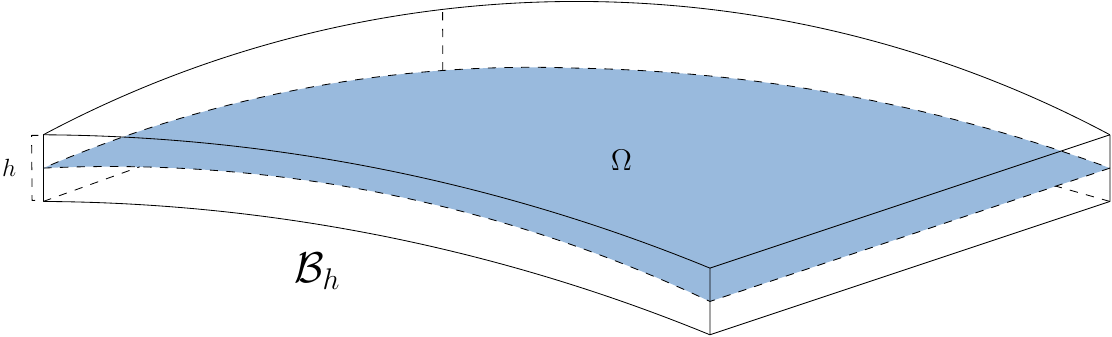}
\caption{}
\end{figure}\label{f:fig1}

As in our previous work \cite{rodriguez2024midsurface}, we employ techniques that have been successful in the settings of classical and fiber-reinforced elastic solids \cite{Koit66, HilgPip92b, HilgPip96, HilgPip97, Steig13, SteigShir19, Steigmann2012fibers, Steigmann2015effects, ShiraniSteigmann2020, steigmann2023lecture} but with special attention paid to the relationship between the intrinsic length scales and $h$. In particular, $\ell_s$ has been estimated in several situations to be much smaller than typical length scales encountered in continuum modeling; see, e.g., \cite{Askes2011, rodriguez2024midsurface, placidi2021granular}. It is then reasonable to assume that
\begin{align}
	\ell_s = O(h), \label{eq:ellsOh}
\end{align}  
for all values of $h > 0$ under consideration. The relation \eqref{eq:ellsOh} implies that the leading order-in-thickness shell energy $U_m$ satisfying 
\begin{align}
	\int_{\mathcal B_h} W(\tilde{\boldsymbol E}, \ell_s\mathrm{Grad}\, \tilde{\boldsymbol E}) \, dV = \int_{\Omega} U_m \, dA + O(h^2), \label{eq:membrane}
\end{align}   
\textit{generally contains nonzero terms of size} $O(h\ell_s^2) = O(h^3)$; see \eqref{eq:surfaceexpansion} and \eqref{eq:W4}. This point cannot be over-emphasized. Similar reasoning also applies to kinetic energies possessing velocity gradients. Thus, \textit{consistent dimensional reductions} starting from three-dimensional kinetic energies possessing velocity gradients and strain-gradient stored energies \textit{must start with cubic order-in-thickness expansions} as in \eqref{eq:cubic}, a central tenet of this work.\footnote{Implicit in the previous discussion is that the stored energy depends in a nontrivial way on the strain-gradients.}  

\subsection{Outline} We now provide a brief outline of this study. Section 2 introduces the geometric and kinematic quantities used to describe the deformation of thin elastic bodies and shells. It also outlines the structure of the stored and kinetic energies, which are assumed to decompose into a sum of a classical contribution (with no additional gradients) and strain-gradient/velocity gradient part. A key assumption is that the intrinsic length scales associated to the strain-gradient/velocity gradient contributions scale with the shell thickness $h$. This implies that the stress vectors on the top and bottom lateral surfaces are approximated up to $O(h^2)$ by those derived from the classical energy contributions alone (see \eqref{eq:PPscomparison}).

Section 3 uses the previous implications for the stress vectors to motivate the choice of director fields in the Taylor expansion (normal to $\Omega$) of the motion, following the methods of Hilgers and Pipkin~\cite{HilgPip96} and Steigmann et al.~\cite{steigmann2023lecture}. Assuming small midsurface strain and a quadratic isotropic classical energy contribution, we show that the stored surface energy satisfies
\begin{align}
U = W_{\mathrm{Koiter}} + W_4 + O(h^4), \label{eq:U}
\end{align}
where $W_4$ incorporates strain-gradient effects from the three-dimensional stored energy.

In Section 4, we derive the explicit form of $W_4$ for isotropic strain-gradient stored energies of both Toupin--Mindlin and dilatational type, completing the explicit computation of the leading cubic order-in-thickness in \eqref{eq:U}. Section 5 then derives the explicit leading cubic order-in-thickness expression for the body's kinetic energy. In both sections, we assume small midsurface strain and a quadratic isotropic classical energy contribution.

Finally, Section 6 illustrates the theory using the simplest strain-gradient model, one derived from dilatational elasticity. We compute the edge tractions and edge double force densities needed to support various equilibrium states including deforming a finite plate into a cylinder, combined extension/compression and radial expansion/contraction of a finite cylinder, and pure torsion of a finite cylinder. 

\bigskip 

\subsection*{Acknowledgements} C. B., R. S. H., and J. M. were supported by NSF RTG DMS-2135998. Y.C. was supported by NSF CAREER Grant DMS-2045494.  C. R. was supported by NSF DMS-2307562. The authors are grateful for this support.

\section{Preliminaries}

In this section, we introduce the geometric and kinematic setup for thin bodies and shells, and we specify the forms of the three-dimensional stored and kinetic energies considered in this work.

\subsection{Kinematics}

We consider a ``thin" elastic piecewise $C^2$ body with reference configuration 
$
	\cl B_h \subseteq \bbE^3
$
where $h > 0$ measures the thickness of the shell, and $\Omega \subset \cl B_h$ is a \textit{base surface} of the body with associated unit normal vector field $\bs N$. We denote a smooth motion of $\cl B_h$ by $$\tilde{\bs \chi} : \cl B_h \times [t_0,t_1] \rar \bbE^3.$$ We assume that $\cl B_h$ can be locally parameterized using \textit{normal coordinate parameterizations}:
\begin{align}
	\bs X(\theta^\alpha, \zeta) = \bs x(\theta^\alpha) + \zeta \bs N(\theta^\alpha) \in \cl B_h,
\end{align} 
where Greek indices take values in $\{1,2\}$, $(\theta^1, \theta^2) \in U \subset \bbR^2$ with $U \subseteq \bbR^2$ an open set, and $\bs x(\theta^\alpha)$ a local parameterization of $\Omega$. Lateral surfaces correspond to constant values for $\zeta \in [-h/2, h/2]$ and the top and bottom lateral surfaces correspond to $\zeta = \pm h/2$, respectively. We follow standard practice and identify $\Omega$ with the \textit{midsurface} of $\cl B_h$. We denote the deformation gradient and its expression as a function of $(\theta^\al, \zeta)$ via 
\begin{align}
	\tilde{\bs F} = \frac{\p \tilde \chi^i}{\p X^a} \bs e_i \tens \bs e^a, \quad \hat{\bs F}(\theta^\al, \zeta) = \tilde{\bs F}(\bs x(\theta^\al) + \zeta \bs N), 
\end{align}  
Here Latin indices take values in $\{1,2,3\}$.
We fix length, time and mass scales $L$, $T$, and $M$ respectively. To decrease notation, we assume that all lengths, times, and masses have been \textit{a priori} non-dimensionalized by $L$, $T$, and $M$ respectively. In particular, we assume that the shell is thin, i.e.,
\begin{align}
h \ll 1.
\end{align}

The natural basis vector fields associated to the local coordinates $(\theta^1, \theta^2)$ are denoted by 
$$\bs A_\alpha = \bs x_{,\alpha}$$
where $\mbox{$\cdot$}_{,\alpha} = \frac{\p}{\p \theta^\al}$. We assume that $\{\bs A_1, \bs A_2, \bs N\}$ form a right-handed basis of $\bbR^3$.  

The components $(A_{\al \beta})$ of the \textit{metric tensor} on $\Omega$ are given by $A_{\al \beta} = \bs A_\al \cdot \bs A_\beta$, and the dual metric components are denoted by $(A^{\al \beta})$. Then  
\begin{align}
	A^{\al \gamma}A_{\gamma \beta} = \delta^\al_\beta. 
\end{align}
The dual basis vector fields are then given by
$\bs A^\alpha = A^{\al \beta} \bs A_\beta$ and satisfy 
\begin{align}
	\bs A_\beta \cdot \bs A^\al = \delta^\al_\beta. 
\end{align}

The \textit{surface gradient} of a scalar, vector or tensor valued function $\bs f$ on $\Omega$ is denoted $\nabla_s \bs f$ and defined via
\begin{align}
	\nabla_s \bs f := \bs f_{,\alpha} \tens \bs A^\alpha. 
\end{align}
The \textit{normal curvature tensor} $\msB$ on the midsurface is the symmetric tensor
\begin{align}
 \msB = -\nabla_s \bs N = (-\bs N_{,\al} \cdot \bs x_{,\beta}) \bs A^\al \tens \bs A^\beta = (\bs N \cdot \bs x_{,\al\beta}) \bs A^\al \tens \bs A^\beta. 
\end{align}
The volume form for $\cl B_h$ can be expressed as
\begin{align}
	dV = \mu_s d\zeta dA, \quad \mu_s := 1-2H \zeta + K\zeta^2, \label{eq:volume}
\end{align}
where $dA$ is the area element on $\Omega$. Here, $H$ and $K$ are the mean and Gauss curvatures, respectively:
\begin{align}
	H = \frac{1}{2} B^\al_\al, \quad K = \det(B^\al_\beta), 
\end{align}
where indices are raised using the dual metric $\bs A^{-1} = A^{\al \beta} \bs A_\al  \tens \bs A_\beta$. We assume that we are working in \emph{shell space}, defined by the condition that $\mu_s > 0$ for all $\zeta \in [-h/2,h/2]$, so that the local parameterizations are orientation preserving. 

\subsection{Assumptions for the three-dimensional kinetic and stored energies}

We assume that the elastic body $\cl B_h$ is homogeneous and the stored energy per unit reference volume is of the strain-gradient form  
\begin{gather}
	W = W_s + W_{sg}, \\
	W_s = \hat W_s(\hat{\bs E}), \quad W_{sg} =   \hat W_{sg}(\hat{\bs E}, \ell_s \mathrm{Grad}\,\hat{\bs E}), \label{eq:stenergy}
\end{gather}
where $\ell_s$ is a characteristic length scale. The subscript ``$s$" denotes a classical strain energy while ``$sg$" denotes a strain-gradient contribution in \eqref{eq:stenergy}. The variable $\hat{\bs E}$ denotes the Lagrange strain as a function of $(\theta^\al, \zeta)$,  
\begin{align}
	\tilde{\bs E} = \frac{1}{2}(\tilde{\bs F}^T \tilde{\bs F} - \bs I), \quad \hat{\bs E}(\theta^\al, \zeta) = \tilde{\bs E}(\bs x(\theta^\al) + \zeta \bs N), 
\end{align} 
and 
\begin{align}
\mathrm{Grad}\, \hat{\bs E}(\theta^\al, \zeta) = \mathrm{Grad}\, \tilde{\bs E}(\bs x(\theta^\al) + \zeta \bs N), \quad \mathrm{Grad}\, \tilde{\bs E} = \Bigl ( \frac{\p}{\p X^a} \tilde{\bs E} \Bigr ) \tens \bs e^a. 
\end{align}
By \eqref{eq:volume},
\begin{align}
	\int_{\cl B_h} W\, dV = \int_\Omega \Bigl (
	\int_{-h/2}^{h/2} W \, \mu_s d\zeta 
	\Bigr ) da,
\end{align}
and in this work, we seek to determine the shell energy $\int_{-h/2}^{h/2} W \, \mu_s d\zeta$ to leading cubic order-in-thickness. 

The reference configuration is the body's natural configuration, and for simplicity, we assume that $W_{sg}$ is quadratic in the strain-gradient
\begin{align}
\begin{split}
	\hat W_s(\bs 0) = 0, \quad \hat W_{s,\hat{\bs E}}(\bs 0) = \bs 0, \\ 
	W_{sg}(\hat{\bs E}, \hat{\bs H}) = \frac{\ell_s^2}{2}\hat{\bs H} \cdot \mathbb D(\hat{\bs E})[\hat{\bs H}], 
	 \label{eq:energyass}
\end{split}
\end{align}
where $\mbox{$\cdot$}_{,\bs A}$ denotes the gradient with respect to the {tensor} variable $\bs A$, and $\hat{\bs H}$ denotes the variable $\mathrm{Grad}\, \hat{\bs E}$. 
Here, $\mathbb{D}(\hat{\bs E})$ takes values in the set of sixth-order tensors that are nonnegative definite
\begin{align}
	\forall \hat{\bs H}, \quad \hat{\bs H} \cdot \mathbb D(\hat{\bs E})[\hat{\bs H}] \geq 0,
\end{align}
and possess symmetries expressed in Cartesian components $\{D_{ijklmn}\}$ via
\begin{align}
	D_{ijklmn} = D_{lmnijk}, \quad D_{ijklmn} = D_{jiklmn} = D_{ijkmln}.
\end{align}
Throughout this work, we assume that the classical contribution to the total stored energy satisfies the strong ellipticity condition for all deformations considered: for all $\bs a \tens \bs b \neq \bs 0$, 
\begin{align}
	(\bs a \tens \bs b)\cdot \cl M_s(\hat{\bs F})[\bs a \tens \bs b] > 0, \label{eq:strongellip}
\end{align}
where
\begin{align}
	\cl M_s := W_{s,\hat{\bs F}\hat{\bs F}}. 
\end{align}

We will also assume that the kinetic energy per unit reference volume $\kappa_R$ of the shell incorporates a velocity gradient contribution,  
\begin{align}
	\kappa_R = \frac{1}{2} \rho_R \Bigl ( |\p_t \hbchi|^2 + \ell_k^2 |\p_t \hbF (\hbF)^{-1}|^2 \Bigr ), \label{eq:kin}
\end{align}
where $\rho_R$ is the constant, reference mass density and $\ell_k$ is a second length scale. In terms of the spatial velocity field $\hbv$, the kinetic energy per unit current volume is given by    
\begin{gather}
	\kappa = \frac{1}{2} \rho \Bigl ( |\bs \hbv|^2 + \ell_k^2 |\mbox{grad}\, \hbv|^2 \Bigr ), \label{eq:kincurr} \\
	\rho = \rho_R (\det \hbF)^{-1}, \quad \mbox{grad}\, \hbv = \frac{\p \hat{v}^i}{\p x^j} \bs e_i \tens \bs e^j.
\end{gather}
By \eqref{eq:volume},
\begin{align}
	\int_{\cl B_h} \kappa_R \, dV = \int_\Omega \Bigl (
	\int_{-h/2}^{h/2} \kappa_R \, \mu_s d\zeta 
	\Bigr ) da, 
\end{align}
and in this work, we seek to determine the kinetic shell energy $\int_{-h/2}^{h/2} \kappa_R \, \mu_s d\zeta$ to leading cubic order-in-thickness.  

One of our two \textbf{fundamental assumptions} is that there exists a dimensionless positive constant $C_1$ such that for all thicknesses under consideration, 
\begin{align}
	\ell_s \leq C_1 h, \quad \ell_k \leq C_1 h. \label{eq:lengthscale}
\end{align}
This assumption reflects the expectation that the length scales $\ell_s$ and $\ell_k$ are comparable to the average interparticle distance or the typical spacing between randomly distributed inhomogeneities in the physical body (being modeled as homogeneous). These are length scales which are not expected to dominate the shell thickness; see \cite{Askes2011, rodriguez2024midsurface, placidi2021granular}.  

We define the following second-order Piola-Kirchhoff tensors $\hbP$ and $\hbP_s$ corresponding to the stored energies $W$ and $W_s$ by 
\begin{gather}
	\hbP = \hat P^{a}_i \bs e^i \tens \bs e_a := \Bigl (\frac{\p W}{\p \hat{F}^i_a} - \frac{\p}{\p X^b} \frac{\p W}{\p \hat{H}^i_{ab} } + \frac{\p}{\p t} \frac{\p \kappa_R}{\p \dot{\hat{F}}^i_{a}} \Bigr ) \bs e^i \tens \bs e_a, \quad 
	\hbP_s := \frac{\p W_s}{\p \hat{F}^i_a} \bs e^i \tens \bs e_a, \label{eq:Pdefinition}
\end{gather}
where $\dot{\hat{F}}^i_{a} = \p_t \hat{F}^i_{a}$. From our expressions for $W$ and $\kappa_R$, we have 
	\begin{align}
		\hbP = \hbP_s + O(\ell_s^2 + \ell_k^2) = \hbP_s + O(h^2),\label{eq:PPscomparison}
	\end{align}
	where the big-oh term depends on the size of $\| \hbchi \|_{C^3(\cl B \times [t_0,t_1])}$.\footnote{We recall that all variables have been nondimensionalized a priori so that \eqref{eq:PPscomparison} is meaningful and $\| \hbchi \|_{C^3(\cl B \times [t_0,t_1])}$ is a dimensionless quantity.} 
As is known (see e.g., \cite{TruesdellNollNLFT} or Section 10 of \cite{Toupin64}), for a body with stored energy $W$ and kinetic energy $\kappa_R$, the traction per unit undeformed area, $\bs t = t_i \bs e^i$, of a surface with unit normal $\hat{\bs N} = \hat N_i \bs e^i$ is given component wise by 
\begin{align}
t_i = \hat P_i^a \hat N_a - \Bigl (D_a \frac{\p W}{\p \hat H^{i}_{ab}} \Bigr ) \hat N_b + \frac{\p W}{\p \hat H^{i}_{ab}}\bigl (\hat B_{ab} - \hat B^a_a \hat N_a \hat N_b \bigr ). \label{eq:traction}
\end{align}
Above, $\hat B_{ab}\bs e^a \tens \bs e^b$ is the normal curvature tensor for the surface, and for a third order tensor $M^{ab}_i \bs e^ \tens \bs e_a \tens \bs e_b$, 
\begin{align}
 D_a M^{ab}_i := M^{ab}_{i,a} - \Bigl ( M^{ab}_{i,c} \hat N^c) \hat N_a.
\end{align}
We note that $\bs t$ is an odd function of the normal vector $\bs N$.

\section{The leading cubic order-in-thickness expressions for the stored shell energy}

In this section, we use the general framework developed by Steigmann and collaborators (see \cite{steigmann2023lecture}) and our length scale assumptions \eqref{eq:lengthscale} to obtain the leading cubic order-in-$h$ expressions for the stored shell energy defined by
\begin{align}
	\int_{-h/2}^{h/2} W \, \mu_s d\zeta.
\end{align}
More precisely, we show that
\begin{align}
	\int_{-h/2}^{h/2} W \, \mu_s d\zeta = W_1 + W_2 + W_3 + W_4 + O(h^4), 
\end{align}
where $W_i$, $i = 1,2,3$, are the same terms arising in the classical elasticity setting; see \eqref{eq:classicalenergies} and \cite{steigmann2023lecture}). The term $W_4$ contains new surface strain-gradient and normal curvature terms. The contribution $W_1 + W_2 + W_3$ yields Koiter's classical shell energy up to an error of size $O(h^4)$ in the case of small midsurface strain and a quadratic isotropic strain energy $W_s$ (see \eqref{eq:quadstrain}). The explicit computation of $W_4$ for the case of a quadratic isotropic strain energy $W_s$ and various forms of $W_{sg}$ is delayed until Section \ref{s:storedenergy}.  

Motions are locally parameterized by
	\begin{align}
		\hbchi = \bs y(\theta^\al, t) + \bs d(\theta^\al, t) \zeta + \bs g(\theta^\al, t) \frac{\zeta^2}{2} + \bs h(\theta^\al, t) \frac{\zeta^3}{6} + O(\zeta^4), \label{eq:chiexp}
	\end{align}
where $\bs y = \hat{\bs \chi} |_{\zeta = 0}$ is the motion of the shell's midsurface $\Omega$, and $\bs d$, $\bs g$, and $\bs h$ are the \textit{directors}. For future use, we record the kinematic formulae 
\begin{gather}
	\bs F = \nabla_s \bs y + \bs d \tens \bs N, \quad 
	\bs F' = \nabla_s \bs d + (\nabla_s \bs y)\bs B + \bs g \tens \bs N. \label{eq:midsurfacedefgrad}
\end{gather}

\subsection{The general method for specifying the directors}

Following Hilgers and Pipkin \cite{HilgPip92b, HilgPip96} and Steigmann et al. \cite{steigmann2023lecture}, we assume that the tractions $\bs t_{\pm}$ on the top and bottom lateral surfaces $\{ \zeta = \pm h/2 \}$ with unit normals $\pm \bs N$, vanish: 
\begin{align}
	\bs t_{\pm} = 0.
\end{align}
Then from \eqref{eq:traction}, our expression for $W$, and \eqref{eq:lengthscale}, we have 
\begin{align}
\bs 0 = \frac{1}{2}(\bs t_{+} - \bs t_-) = \bs P \bs N + O(h^2), \quad \bs 0 = \frac{1}{2}(\bs t_+ + \bs t_-) = h\bs P' \bs N + O(h^3),
\end{align} 
where the absence of a caret indicates evaluation at $\zeta = 0$ and $\mbox{}' = \frac{d}{d\zeta} |_{\zeta = 0}$. Then by \eqref{eq:PPscomparison}, 
\begin{align}
	\bs P_s \bs N = O(h^2), \quad \bs P'_s \bs N = O(h^2). \label{eq:midpiola} 
\end{align}
Motivated by \eqref{eq:midpiola} and \cite{HilgPip92b, HilgPip96, steigmann2023lecture}, our \textbf{second fundamental assumption} is that 
\begin{align}
	\bs P_s \bs N = \bs 0, \quad \bs P'_s \bs N = \bs 0. \label{eq:steigcond}
\end{align}
As shown by Hilgers and Pipkin \cite{HilgPip92a, HilgPip96}, the strong ellipticity assumption on $W_s$, \eqref{eq:strongellip}, implies that \eqref{eq:steigcond} has unique solutions $\bs d = \bar{\bs d}$, $\bs g = \bar{\bs g}$, and
\begin{align}
	\bs P_s(\nabla_s \bs y + \bar{\bs d} \tens \bs N)\bs N &= \bs 0, \label{eq:choiced} \\
	\cl A_s(\nabla_s \bs y + \bar{\bs d} \tens \bs N)\bar{\bs g} &= - 
	\cl M_s(\nabla_s \bs y + \bar{\bs d} \tens \bs N)[\nabla_s \bar{\bs d} + (\nabla_s \bs y)\bs B] \bs N, \label{eq:choiceg}
\end{align}
where $\bs P_s(\cdot)$ is the Piola-Kirchoff tensor evaluated at $\zeta = 0$ and $\cl A_s$ is the acoustic tensor defined by
\begin{align}
	\cl A_s(\hat{\bs F})\bs u = \bigl (\cl M_s(\hat{\bs F})[\bs u \tens \bs N]\bigr )\bs N. 
\end{align}
We denote the kinematic variables and Piola-Kirchhoff stress associated to these choices of directors with an overbar: 
\begin{gather}
	\bar{\bs F} = \nabla_s \bs y + \bar{\bs d} \tens \bs N, \quad \bar{\bs F}' = \nabla_s \bar{\bs d} + (\nabla_s \bs y) \bs B + \bar{\bs g} \tens \bs N, \label{eq:Fprime}\\
	\bar{\bs P}_s = \bs P_s(\bar{\bs F}). 
\end{gather}

\subsection{The leading cubic-in-$h$ order expression for the stored shell energy}

The previous choices of $\bs d$ and $\bs g$ determine the leading cubic order-in-$h$ expression for the stored shell energy corresponding to the classical contribution $W_s$. By Sections 4-5 of \cite{Steig13} (see also Chapter 6 of \cite{steigmann2023lecture}), it follows that 
\begin{align}
	\int_{-h/2}^{h/2} W_s \, \mu_s d\zeta = 
	\Bigl (1 + \frac{1}{12}h^2 K \Bigr ) W_1 + W_2 + W_3 + O(h^4), \label{eq:cubicclassical}
\end{align}
where 
\begin{gather}
	W_1 = h W_s(\nabla_s \bs y + \bar{\bs d}\tens \bs N), \\ \quad W_2 = \frac{h^3}{24} \bs K \cdot \cl M_s(\nabla_s \bs y + \bar{\bs d} \tens \bs N)[\bs K], \quad \bs K := \nabla_s \bar{\bs d} + (\nabla_s \bs y) \bs B + \bar{\bs g} \tens \bs N, \\
	W_3 = \frac{h^3}{12} \bar{\bs P}_s\bs 1 \cdot \bigl \{
	[(\nabla_s \bar{\bs d}) \bs B + (\nabla_s \bs y) \bs B^2] - 2 H[ \nabla_s \bar{\bs d} + (\nabla_s \bs y)\bs B]
	\bigr \}. \label{eq:classicalenergies}
\end{gather}
The leading cubic order-in-$h$ contribution to the stored shell energy is now readily obtained. By \eqref{eq:stenergy}, \eqref{eq:cubicclassical}, \eqref{eq:energyass}, \eqref{eq:lengthscale}, 
\begin{align}
	\int_{-h/2}^{h/2} W \, \mu_s d\zeta =
	\Bigl (1 + \frac{1}{12} h^2 K\Bigr )W_1 + W_2 + W_3 + W_4 + O(h^4), \label{eq:surfaceexpansion}
\end{align}
where 
\begin{gather}
	W_4 = \frac{1}{2} h \ell_s^2 \bar{\bs H} \cdot \bbD(\bs 0)[\bar{\bs H}], \label{eq:W4}\\
	\bar{\bs H} := \mathrm{Grad}\,\bar{\bs E} = \nabla_s \bar{\bs E} + \bar{\bs E}' \tens \bs N, \quad
	\bar{\bs E} = \frac{1}{2}(\bar{\bs F}^T\bar{\bs F} - \bs I), \quad \bar{\bs E}' = \mathrm{sym}\bigl (\bar{\bs F}^T\bar{\bs F}' \bigr ). \qquad \label{eq:barH}
\end{gather}

\subsection{The choice of the directors when $W_s$ is a quadratic isotropic strain energy}\label{s:quad}

In this section, we recall the form of $\bar{\bs d}$ and $\bar{\bs g}$ derived for 
\begin{align}
W_s = \frac{\la}{2}(\tr \hat{\bs E})^2 + \mu|\hat{\bs E}|^2, \label{eq:quadstrain}
\end{align}
where $\lambda$ and $\mu$ are the classical Lam\'e parameters of the material and
assuming the midsurface strain satisfies,
\begin{align}
	|\bs E| \leq C_2 h, \label{eq:smallstrain}
\end{align}
with $C_2$ a positive constant. See Section 6.1.7 and 6.2 of \cite{steigmann2023lecture} for detailed derivations.\footnote{We recall that all variables have been nondimensionalized a priori so that $C_2$ is a dimensionless constant.}
 
The assumption \eqref{eq:smallstrain} implies that ${\bs F} = \bs R + O(h)$, where $\bs R$ is a rotation valued field on $\Omega$, and up to an $O(h)$ error,  
\begin{align}
	\forall \alpha = 1, 2, \quad \bs y_{,\al} = \bs R \bs A_\al, \quad \nabla_s \bs y = \bs R \bs 1, \quad 
	\bs 1 = \bs A_\al \tens \bs A^\al, \quad \bs n = \bs R \bs N. \label{eq:smallstrainconsequences}
\end{align} 
We will then conclude this section by using the form of $\bar{\bs d}$ and $\bar{\bs g}$ to obtain the form of the strain-gradient when restricted to the midsurface, given by $\eqref{eq:barH}_1$.

We denote the following components
\begin{align}
	 \hat E_{\al 3} := \bs A_\al \cdot \hat {\bs E} \bs N \quad \mbox{and} \quad \hat E_{33} = \bs N \cdot \hat {\bs E} \bs N.
\end{align}
Then $\bs 0 = \bs P_s \bs N = \bs F \bs S \bs N$ with
\begin{align}
\bs S := (W_{s})_{\hat{\bs E}} = \la (\tr \hat{\bs E}) \bs I + 2\mu \hat{\bs E}
\end{align}
implies that
\begin{align}
\forall \al, \beta = 1,2, \quad \bar E_{\al 3} = 0, \quad \bar E_{33} = -\frac{\lambda}{\lambda + 2\mu} \tr \bs \eps,  \label{eq:E33}
\end{align}
where $\bs \eps = \bar{E}_{\al\beta} \bs A^\al \tens \bs A^\beta = \frac{1}{2}(a_{\al \beta} - A_{\al \beta}) \bs A^\al \tens \bs A^\beta$, and 
$$a_{\al \beta} = \bs y_{,\al} \cdot \bs y_{,\beta}$$ are the components of the \textit{metric tensor on the current configuration of the midsurface}. Thus, 
\begin{gather}
	\bar{\bs E} = \bs \eps + \bar E_{33} \bs N \tens \bs N, \label{eq:barE} \\
	\bar{\bs d} = \varphi \bs n, \quad \varphi = (1 + 2 \bar E_{33} )^{1/2}. \label{eq:bard2}
\end{gather}
In particular, 
\begin{align}
	\varphi = 1 -\frac{\lambda}{\lambda + 2\mu} \tr \bs \eps + O(h^2), \label{eq:varphi}
\end{align}
and 
\begin{align}
	\nabla_s \bar{\bs d} = -\bs R \bs \kappa + \bs R \bs N \tens \nabla_s \varphi + O(h),
\end{align}
where
\begin{align}
	\bs \kappa = -b_{\al \beta} \bs A^\al \otimes \bs A^\beta, \quad b_{\al \beta} = -\bs n_{,\al} \cdot \bs y_{,\beta} = \bs n \cdot \bs y_{,\al\beta},
\end{align}
see (6.181) in \cite{steigmann2023lecture}. We note that $\bs \kappa$ is (up to a sign) the pullback of \textit{the normal curvature tensor on the current configuration} of the midsurface. Using \eqref{eq:smallstrainconsequences}, it follows that 
\begin{gather}
	\bs R^T (\nabla_s \bar{\bs d} + (\nabla_s \bs y)\bs B) = - \bs \rho + \bs N \tens \nabla_s \varphi + O(h), \label{eq:bard}
\end{gather}
where 
$$\bs \rho = -(\bs \kappa + \bs B) = (b_{\al \beta} - B_{\al \beta}) \bs A^\al \tens \bs A^\beta.$$

The form of $\bs R^T \bar{\bs g}$ is derived from the following two relations (see (6.178) from \cite{steigmann2023lecture}) that are valid up to errors of size $O(h)$:
\begin{align}
	\bs N \cdot \bs R^T \bar{\bs g} &= -(\lambda + 2\mu)^{-1} \Bigl \{ \la
	\tr(\bs R^T \nabla_s \bar{\bs d} + \bs B) + 2 \mu \bs N \cdot \bigl [\mathrm{sym}(\bs R^T \nabla_s \bar{\bs d} + \bs B)\bigr ]\bs N
	\Bigr \}, \\
	\bs 1(\bs R^T \bar{\bs g}) &= -2 \bs 1 \bigl [\mathrm{sym}(\bs R^T \nabla_s \bar{\bs d} + \bs B)\bigr ]\bs N.
\end{align}
Then $\bs R^T \bar{\bs g} = \bs 1(\bs R^T \bar{\bs g}) + (\bs N \cdot \bs R^T \bar{\bs g}) \bs N$ and \eqref{eq:bard} imply that 
\begin{align}
	\bs R^T \bar{\bs g} = \frac{\lambda}{\lambda + 2\mu}(\tr \bs \rho)\bs N - \nabla_s \varphi + O(h). \label{eq:barg}
\end{align}

The assumption \eqref{eq:smallstrain}, \eqref{eq:varphi}, \eqref{eq:bard}, and \eqref{eq:barg} imply that up to $O(h^4)$ errors that don't affect the leading cubic-in-$h$ order behavior of $\int_{-h/2}^{h/2} W \, \mu d\zeta$, we have    
\begin{align}
	W_1 + W_2 + W_3 &= h \Bigl (
	\frac{\lambda \mu}{\lambda + 2\mu}(\tr \bs \eps)^2 + \mu |\bs \eps|^2
	\Bigr ) + \frac{h^3}{24} \Bigl ( \frac{\lambda \mu}{\lambda + 2\mu}(\tr \bs \rho)^2 + \mu |\bs \rho|^2 \Bigr )\\
	&=: W_{\mathrm{Koiter}}. \label{eq:Koiter}
\end{align} 
The energy \eqref{eq:Koiter} is Koiter's classical shell energy first derived in \cite{Koit66}. By \eqref{eq:surfaceexpansion}, it follows that 
\begin{align}
	\int_{-h/2}^{h/2}W \, \mu_s d\zeta = W_{\mathrm{Koiter}} + W_4 + O(h^4), \label{eq:Wexpansion}
\end{align} 
where $W_4$ is defined in \eqref{eq:W4} and \eqref{eq:barH}. Moreover, the $O(h^4)$ term depends on $C_1$ appearing in \eqref{eq:lengthscale}, $C_2$ appearing in \eqref{eq:smallstrain}, and $\| \hat{\bs \chi} \|_{C^3(\cl B_h \times [t_0, t_1])}$. 
 
We now derive the form of the strain-gradient restricted to the midsurface (see \eqref{eq:barH}) associated with our choice of \( \bar{\bs d} \) and \( \bar{\bs g} \) for the energy \eqref{eq:quadstrain}. By $\eqref{eq:Fprime}_2$, \eqref{eq:bard}, \eqref{eq:barg}, and the small midsurface strain assumption, we obtain 
\begin{align}
\mathrm{sym}(\bar{\bs F}^T \bar{\bs F}') &= \mathrm{sym}\bigl (
\bs R^T \nabla_s \bar{\bs d} + \bs R^T(\nabla_s \bs y)\bs B + \bs R^T\bar{\bs g} \tens \bs N 
\bigr )  + O(h) \\
&= \mathrm{sym}\Bigl (
- \bs \rho + \bs N \tens \nabla_s \varphi + \frac{\lambda}{\lambda+2\mu}(\tr \bs \rho) \bs N \tens \bs N \\
&\qquad- \nabla_s \varphi \tens \bs N
\Bigr )  + O(h) \\
&= -\bs \rho + \frac{\lambda}{\lambda+2\mu}(\tr \bs \rho) \bs N \tens \bs N + O(h). \label{eq:barFTF}
\end{align}
By \eqref{eq:E33}, \eqref{eq:barE}, \eqref{eq:barH}, and \eqref{eq:barFTF}, we conclude that
\begin{align}
\begin{split}
	\bar{\bs H} := \mathrm{Grad}\ \bar{\bs E} &= \nabla_s \bs \eps - \frac{\lambda}{\lambda+2\mu} \bs N \tens \bs N \tens \nabla_s(\tr \bs \eps)\\
	&\quad -\bs \rho \tens \bs N + \frac{\lambda}{\lambda+2\mu}(\tr \bs \rho) \bs N \tens \bs N \tens \bs N + O(h). 
\end{split}\label{eq:surfsg}
\end{align}
In Section \ref{s:storedenergy}, we will use \eqref{eq:surfsg} to compute the strain-gradient contribution \( W_4 \) (see \eqref{eq:W4}, \eqref{eq:barH}) to the shell energy expansion \eqref{eq:surfaceexpansion}, assuming \( W_s \) is given by \eqref{eq:quadstrain} and \( W_{sg} \) belongs to a broad class of strain-gradient energy densities.  

\section{Stored shell energies derived from three-dimensional isotropic strain-gradient stored energies}\label{s:storedenergy}

In this section, we complete the derivation of a class of strain-gradient shell energies that arise as the leading cubic order-in-\( h \) contribution to the integral \( \int_{-h/2}^{h/2} W\, \mu_s\, d\zeta \), assuming \( W_s \) is given by \eqref{eq:quadstrain}. As in the previous section, we consider deformations with midsurface strain satisfying,
\begin{align}
	|\bs E| \leq C_2 h, \quad \text{on } \Omega, \label{eq:smallmidsurfacestrain}
\end{align}
where $C_2$ is a positive constant.\footnote{We recall that all variables have been nondimensionalized a priori so that $C_2$ is dimensionless.} We derive two-dimensional shell energies from the following three-dimensional isotropic strain-gradient stored energies:
\begin{itemize}
\item Strain-gradient energies of Toupin-Mindlin type \cite{Toupin62, Toupin64, Mindlin64a, MindlinEshel1968}\footnote{The inner product of two $3$-tensors $\bs A$ and $\bs B$ is given in Cartesian coordinates by $\bs A \cdot \bs B := \sum_{ijk=1}^3 A_{ijk} B_{ijk}$. The norm $|\bs A|$ of a $3$-tensor $\bs A$ is then defined via $|\bs A|^2 = \bs A \cdot \bs A.$}:  
\begin{align}
\begin{split}
	W &= \frac{\la}{2} (\tr \hat{\bs E})^2 + \mu |\hat{\bs E}|^2 + a_1 \mathrm{Grad}(\tr\hat{\bs E}) \cdot \mathrm{Div}\, \hat{\bs E} + a_2 |\mathrm{Grad}(\tr\hat{\bs E})|^2 \\&+ a_3 |\mathrm{Div}\, \hat{\bs E}|^2 +
	a_4 |\mathrm{Grad}\,\hat{\bs E}|^2 + a_5 \mathrm{Grad}\,\hat{\bs E}\cdot \mathrm{Grad}\,\hat{\bs E}^{T(2,3)}.
	\end{split}\label{eq:mindesh}
\end{align}
where $T(i,j)$ denotes taking the transpose in the $i$th and $j$th indices. Strong ellipticity of the associated infinitesimal theory derived from \eqref{eq:mindesh} wherein $\hat{\bs E}$ is replaced by the infinitesimal strain tensor requires that 
\begin{align}
	\sum_{j = 1}^5 a_j > 0 \quad \mbox{and} \quad a_3 + 2a_4 + a_5 > 0;  \label{eq:strongellipticity}
\end{align}
see \cite{eremeyev2022strong}. Moreover, the third through seventh terms in \eqref{eq:mindesh} form the most general expression for an isotropic quadratic form of the strain-gradients. By \eqref{eq:strongellipticity}, there are two length scales associated to the energy \eqref{eq:mindesh},
\begin{align}
	\ell_1^2 := (\lambda + 2\mu)^{-1}\sum_{j = 1}^5 a_j, \quad \ell_2^2 := (2\mu)^{-1}(a_3 + 2 a_4 + a_5).
\end{align}
The recent works \cite{placidi2021granular, placidi2025granular} offer compelling micromechanical arguments for granular materials that $a_1, \ldots, a_5$ are all proportional to $L^2$ where $L$ is the average distance between nearest-neighbor grains, and the proportionality constants depend only on the classical Lam\'e parameters of the material. Motivated by their work, we assume that $a_1, \ldots, a_5$ are all proportional to a single length scale squared $\ell_s^2$ satisfying $\ell_s \leq C_1 h$.  
 
\item Dilatational strain-gradient energies \cite{eremeyev2021nonlinear}: 
\begin{align}
	W &= \frac{\la}{2} (\tr \hat{\bs E})^2 + \mu |\hat{\bs E}|^2 + \frac{1}{2}{\ell_s^2 \mu}\gamma(\hat{\bs E})|\mathrm{Grad}(\det \hat{\bs F})|^2, \label{eq:dilat}
\end{align}
where $\gamma$ is a positive scalar valued function with $\gamma(\bs 0) = 1$ and $\ell_s \leq C_1 h$. A dilatational model may be interpreted as a capillary fluid (originally introduced in \cite{korteweg1901}) undergoing ``solidification"; see Section 3 of \cite{eremeyev2021nonlinear}.

\end{itemize}
In both cases, we have that 
\begin{align}
	W_s = \frac{\la}{2} (\tr \hat{\bs E})^2 + \mu |\hat{\bs E}|^2. 
\end{align}
In the sequel, we will use the results from Section \ref{s:quad} that the strain-gradient restricted to the midsurface corresponding to $\bar{\bs d}, \bar{\bs g}$ is given by
\begin{align}
\begin{split}
\bar{\bs H} := \mathrm{Grad}\, \bar{\bs E} &= \nabla_s \bs \eps - \frac{\lambda}{\lambda+2\mu} \bs N \tens \bs N \tens \nabla_s(\tr \bs \eps)\\
&\quad -\bs \rho \tens \bs N + \frac{\lambda}{\lambda+2\mu}(\tr \bs \rho) \bs N \tens \bs N \tens \bs N + O(h). 
\end{split}\label{eq:straingradient}
\end{align}
In particular, we make use of the identities 
\begin{gather}
\mathrm{Grad}(\tr \bar{\bs E}) = \tr_{(1,2)}\bar{\bs H}
= \frac{2\mu}{\lambda + 2\mu}\bigl (\nabla_s (\mathrm{tr}\, \bs \eps) - (\tr \bs \rho) \bs N \bigr ), \label{eq:gradtrace} \\
\mathrm{Div}\,\bar{\bs E} = \tr_{(2,3)}\bar{\bs H} = \mathrm{div}_s\bs \eps + \frac{\lambda}{\lambda + 2\mu}(\tr \bs \rho) \bs N, \label{eq:divergence}
\end{gather}
where $\tr_{(i,j)}$ denotes taking the trace in the $i$th and $j$th indices. 

\subsection{Toupin-Mindlin strain-gradient energies}
Since \eqref{eq:mindesh} is quadratic, it follows using \eqref{eq:gradtrace} and \eqref{eq:divergence} that
\begin{gather}
	W_4 = ha_1 \tr_{(1,2)}\bar{\bs H} \cdot \tr_{(2,3)}\bar{\bs H} + h a_2 |\tr_{(1,2)}\bar{\bs H}|^2 + ha_3 |\tr_{(2,3)}\bar{\bs H}|^2 \\+ ha_4 |\bar{\bs H}|^2 + ha_5 \bar{\bs H} \cdot \bar{\bs H}^{T(2,3)} + O(h^4). 
\end{gather}
Using \eqref{eq:straingradient}, \eqref{eq:gradtrace}, and \eqref{eq:divergence} we conclude that 
\begin{align*}
	W_4 
        &= h a_1 \Bigl [
	\frac{2\mu}{\lambda + 2\mu} (\nabla_s \tr \bs \eps) \cdot \mathrm{div}_s \bs \eps - \frac{2\lambda \mu}{(\lambda+2\mu)^2} (\tr \bs \rho)^2 
	\Bigr ] \\ 
        &+ h a_2 \frac{4\mu^2}{(\lambda + 2 \mu)^2}\bigl [|\nabla_s \tr \bs \eps|^2 + (\tr \bs \rho)^2] \\
	&+ h a_3 \Bigl [ |\mathrm{div}_s \bs \eps|^2 + \frac{\la^2}{(\la + 2\mu)^2} (\tr \bs \rho)^2 \Bigr ] \\
	&+ h a_4 \Bigl [
	|\nabla_s \bs \eps|^2 + \frac{\la^2}{(\la + 2\mu)^2} |\nabla_s \tr \bs \eps|^2 + |\bs \rho|^2 + \frac{\la^2}{(\la + 2\mu)^2} |\tr \bs \rho|^2
	\Bigr ] \\
	&+ h a_5 \Bigl [
	(\nabla_s \bs \eps) \cdot (\nabla_s \bs \eps)^{T(2,3)} + \frac{\la^2}{(\la + 2 \mu)^2} (\tr \bs \rho)^2 
	\Bigr ] + O(h^4). 
\end{align*}
In summary, we have for \eqref{eq:mindesh}
\begin{align*}
	\int_{-h/2}^{h/2} W \, \mu_s d\zeta 
        &= W_{\mathrm{Koiter}} + ha_1 \Bigl [
	\frac{2\mu}{\lambda + 2\mu} (\nabla_s \tr \bs \eps) \cdot \mathrm{div}_s \bs \eps - \frac{2\lambda \mu}{(\lambda+2\mu)^2} (\tr \bs \rho)^2 
	\Bigr ] \\
        &+ ha_2 \frac{4\mu^2}{(\lambda + 2 \mu)^2}\bigl [|\nabla_s \tr \bs \eps|^2 + (\tr \bs \rho)^2] \\
	&+ ha_3 \Bigl [ |\mathrm{div}_s \bs \eps|^2 + \frac{\la^2}{(\la + 2\mu)^2} (\tr \bs \rho)^2 \Bigr ] \\
	&+ ha_4 \Bigl [
	|\nabla_s \bs \eps|^2 + \frac{\la^2}{(\la + 2\mu)^2} |\nabla_s \tr \bs \eps|^2 + |\bs \rho|^2 + \frac{\la^2}{(\la + 2\mu)^2} |\tr \bs \rho|^2
	\Bigr ] \\
	&+ ha_5 \Bigl [
	(\nabla_s \bs \eps) \cdot (\nabla_s \bs \eps)^{T(2,3)} + \frac{\la^2}{(\la + 2 \mu)^2} (\tr \bs \rho)^2 
	\Bigr ] + O(h^4), \label{eq:toupinmindlin}
\end{align*}
where the $O(h^4)$ depends on $C_1$ appearing in \eqref{eq:lengthscale}, $C_2$ appearing in \eqref{eq:smallstrain}, and $\| \hat{\bs \chi} \|_{C^3(\cl B_h)}$. 

\subsection{Dilatational strain-gradient energies}

We write $$\det \bs F = [\det(\bs I + 2\hat{\bs E})]^{1/2}$$ and
\begin{align}
W_{sg} = \frac{1}{2}{\mu \ell_s^2}\gamma(\hat{\bs E}) \bigr |\mathrm{Grad}[\det(\bs I + 2\hat{\bs E})]^{1/2}\bigr |^2. 
\end{align}
Now 
\begin{align}
\mathrm{Grad}[\det(\bs I + 2\hat{\bs E})]^{1/2} &= [\det(\bs I + 2\hat{\bs E})]^{1/2}
\tr_{(1,2)} \Bigl [ (\bs I + 2 \hat{\bs E})^{-1/2} \mathrm{Grad}\, \hat{\bs E} 
\Bigr ] \\
&= \tr_{(1,2)}\mathrm{Grad}\, \hat{\bs E} + O(h)
\end{align}
and thus, by \eqref{eq:lengthscale} 
\begin{align}
W_4 = \frac{1}{2}{h\ell_s^2 \mu} |\tr_{(1,2)}\bar{\bs H}|^2 + O(h^4).
\end{align}
By \eqref{eq:gradtrace}, we conclude that
\begin{align}
	W_4 &= \frac{1}{2}{h\ell_s^2 \mu}\Bigl |\frac{2\mu}{\lambda + 2\mu} \nabla_s (\tr \bs \eps) - \frac{2\mu}{\lambda + 2\mu}(\tr \bs \rho) \bs N \Bigr |^2 + O(h^4)\\
	&= {h\ell_s^2 \mu} \frac{2 \mu^2}{(\lambda + 2\mu)^2}\bigl ( |\nabla_s (\tr \bs \eps)|^2 + |\tr \bs \rho|^2 \bigr ) + O(h^4). 
\end{align}
In summary, if $W$ is given by \eqref{eq:dilat}, then  
\begin{align}
\int_{-h/2}^{h/2} W \, \mu_s d\zeta = W_{\mathrm{Koiter}} + {h\ell_s^2 \mu}\frac{2 \mu^2}{(\lambda + 2\mu)^2}\bigl ( |\nabla_s (\tr \bs \eps)|^2 + |\tr \bs \rho|^2 \bigr ) + O(h^4), \label{eq:dilatsurfaceenergy}
\end{align}
where the $O(h^4)$ depends on $C_1$ appearing in \eqref{eq:lengthscale}, $C_2$ appearing in \eqref{eq:smallstrain}, and $\| \hat{\bs \chi} \|_{C^3(\cl B_h)}$.

\section{Kinetic shell energies derived from three-dimensional kinetic energies with velocity gradients}
In this section, we derive a class of shell kinetic energies that arise as the leading cubic order-in-\( h \) contribution to the integral  
\begin{align}
	\int_{-h/2}^{h/2} \kappa_R \, \mu_s d\zeta, 
\end{align}
where $\kappa_R$ is given in \eqref{eq:kin}. An explicit expression is obtained when $W_s$ is given by \eqref{eq:quadstrain}. 
In addition to the small midsurface strain assumption \eqref{eq:smallstrain}, we assume that on $\Omega \times [t_0,t_1]$,\footnote{We recall that all variables have been nondimensionalized a priori so that $C_3$ is a dimensionless constant.}
\begin{align}
	|\mathrm{Grad} \, \p_t {\bs F}[\bs N \tens \bs N] \bigr | \leq C_3 h, \label{eq:kinassumpt}
\end{align}
or equivalently, on $\Omega \times [t_0,t_1]$, 
\begin{align}
	|\p_t \bs g| \leq C_3 h. \label{eq:kinassum}
\end{align}

We have by \eqref{eq:kin}, \eqref{eq:chiexp}, \eqref{eq:lengthscale}, and \eqref{eq:kinassum} that the kinetic energy satisfies  
\begin{align*}
	\int_{-h/2}^{h/2} \kappa_R \, \mu_s d\zeta 
        &= \int_{-h/2}^{h/2} \frac{1}{2} \rho_R \Bigl (
	|\p_t \hbchi|^2 + \ell_k^2 |\p_t \hbF (\hbF)^{-1}|^2 
	\Bigr ) \mu_s d\zeta \\
        &= \frac{1}{2}\rho_s \Bigl [ \Bigl (1 + \frac{h^2}{12}K \Bigr )
	|\p_t \bs y|^2 + \frac{h^2}{12} |\p_t \bs d|^2 - \frac{h^2}{3} H \p_t \bs y \cdot \p_t \bs d \\
	&\quad+ \ell_k^2 |\p_t \bs F (\bs F)^{-1}|^2 + \frac{h^2}{12}
	\p_t \bs y \cdot \p_t \bs g \Bigr ] + O(\ell_k^2 h^3) + O(h^5), \\
	&=  \frac{1}{2}\rho_s \Bigl [ \Bigl (1 + \frac{h^2}{12}K \Bigr )
	|\p_t \bs y|^2 + \frac{h^2}{12} |\p_t \bs d|^2 - \frac{h^2}{3} H \p_t \bs y \cdot \p_t \bs d \\
        &\quad+ \ell_k^2 |\p_t \bs F (\bs F)^{-1}|^2 \Bigr ] + O(h^4),
	\label{eq:kineticexpproof}
\end{align*}
where $\rho_s = h \rho_R$ is the mass per unit reference area of the midsurface. Again, by \eqref{eq:smallstrain} we have 
$\bs F = \bs R + O(h)$ where $\bs R$ takes values in the group of rotations, and thus, by \eqref{eq:midsurfacedefgrad} we have  
\begin{gather}
	|\p_t \bs F (\bs F)^{-1}|^2 = |\p_t \bs F \bs R^T|^2 + O(h) = |\p_t \bs F|^2 + O(h) \\
	= |\nabla_s \p_t \bs y|^2 + |\p_t \bs d|^2 + O(h). \label{eq:gradkinsimpl}
\end{gather}
From \eqref{eq:kineticexpproof}, \eqref{eq:gradkinsimpl}, and \eqref{eq:lengthscale}, we conclude that 
\begin{gather}
	\int_{-h/2}^{h/2} \kappa_R \, \mu_s d\zeta = \frac{1}{2}\rho_s \Bigl [ \Bigl (1 + \frac{h^2}{12}K \Bigr )
	|\p_t \bs y|^2 + \frac{h^2}{12} |\p_t \bs d|^2 - \frac{h^2}{3} H \p_t \bs y \cdot \p_t \bs d \\
	+ \ell_k^2 |\nabla_s \p_t \bs y|^2 + \ell_k^2 |\p_t \bs d|^2\Bigr ] + O(h^4). \label{eq:kineticexpproof2}
\end{gather}
It is straightforward to see that \eqref{eq:kineticexpproof2} is positive definite in $\p_t \bs y$ and $\p_t \bs d$ for all values of $\ell_k$ if and only if 
\begin{align}
	h^2(4 H^2 - K) < 12. \label{eq:posdefkin}
\end{align}

As a special case, suppose that $W_s$ is given by \eqref{eq:quadstrain} and $\bs d = \bar{\bs d}$ with $\bar{\bs d}$ given by \eqref{eq:bard2}. Then 
\begin{align}
	\p_t \bs d = -\frac{\la}{\la+2\mu}(\tr \p_t \bs \eps) \bs n + \p_t \bs n + O(h),
\end{align}
and since $\bs n \cdot \p_t \bs n = 0$,  
\begin{align*}
	\int_{-h/2}^{h/2} \kappa_R \, \mu_s d\zeta 
        &= \frac{1}{2}\rho_s \Bigl [ \Bigl (1 + \frac{h^2}{12}K \Bigr )
	|\p_t \bs y|^2 + \frac{h^2}{12} \frac{\la^2}{(\la + 2\mu)^2} (\tr \p_t \bs \eps)^2 + \frac{h^2}{12}|\p_t \bs n|^2 \\
	&\qquad\qquad + \frac{h^2}{3} H \p_t \bs y \cdot \Bigl (\frac{\la}{\la+2\mu}(\tr \p_t \bs \eps) \bs n - \p_t \bs n\Bigr ) \\
	&\qquad\qquad+ \ell_k^2 |\nabla_s \p_t \bs y|^2 + \ell_k^2 \frac{\la^2}{(\la + 2\mu)^2} (\tr \p_t \bs \eps)^2 + \ell_k^2 |\p_t \bs n|^2 \Bigr ] + O(h^4). \label{eq:kineticexp}
\end{align*}
The right-hand side of \eqref{eq:kineticexp} represents the leading cubic order-in-$h$ contribution to the kinetic energy in the case that $W_s$ is given by \eqref{eq:quadstrain}, assuming \eqref{eq:smallstrain} and \eqref{eq:kinassumpt}. It is positive definite in $\p_t \bs y$, $\frac{\lambda}{\lambda + 2\mu}\tr \p_t \bs \eps$, and $\p_t \bs n$ if and only if \eqref{eq:posdefkin} holds. The $O(h^4)$ term depends on $C_1$ appearing in \eqref{eq:lengthscale}, $C_2$ appearing in \eqref{eq:smallstrain}, $C_3$ appearing in \eqref{eq:kinassumpt}, and $\| \hat{\bs \chi} \|_{C^4(\cl B_h \times [t_0,t_1])}$.\footnote{We recall that all quantities have been nondimensionalized a priori, and thus, $\| \hat{\bs \chi} \|_{C^4(\cl B_h \times [t_0,t_1])}$ is a dimensionless constant.}

\section{Equilibrium states of a finite cylinder}

In this section, we consider a cylinder of radius $R$ and length $L$: 
\begin{align}
	\Omega = \Bigl \{ (R \cos \theta, R \sin \theta, z) \mid (\theta, z) \in [-\pi, \pi] \times [0,L] \Bigr \}.   
\end{align}
We will determine the body force $\bs g$, edge tractions $\bs t$, and edge double force densities $\bs m$ necessary to support several equilibrium states including a finite plate deformed into a cylinder, combined extension/compression and radial expansion/contraction of a finite cylinder, and pure torsion of a finite cylinder. In our calculations, for simplicity we will assume that the shell energy $\scrW$ is given by the leading cubic-in-order expression appearing in \eqref{eq:dilatsurfaceenergy},
\begin{gather}
\scrW = W_{\mathrm{Koiter}} + {h\ell_s^2 \mu}\frac{2 \mu^2}{(\lambda + 2\mu)^2}\bigl ( |\nabla_s (\tr \bs \eps)|^2 + |\tr \bs \rho|^2 \bigr )
\end{gather}  

\subsection{General equilibrium equations}

We first record the general equilibrium equations for shells satisfying the virtual work equality
\begin{align}
    \dot{E} = P,
\end{align}
where the dot denotes the variational derivative, $P$ represents the virtual work, and $E$ is the stored energy of the shell, given by
\begin{align}
    E = \int_\Omega \scrW \, dA.
\end{align}
The shell is assumed to have piecewise smooth boundary with finitely many corners indexed by $i \in \{1,\ldots,N\}$. Our derivation largely follows that of \cite{Steigmann18Lattice}. 

Let $\bs u = \dot{\bs{y}}$, the variational derivative of our position field $\bs{y}$ on the deformed configuration $\omega$. We define $S^\mu_{\alpha \beta} = \Gamma^\mu_{\alpha \beta} - \overline{\Gamma}^\mu _{\alpha \beta}$, which denote the differences in Christoffel symbols on the current surface, $\omega$, and the reference surface, $\Omega$.
We have that 
\begin{equation}\label{eq:storedenergyvariation}
    \dot{\scrW} = \bs{N}^\alpha\cdot \bs{u}_{,\alpha} + \bs{M}^{\alpha \beta} \cdot \bs{u}_{|\alpha \beta},
\end{equation} where 
\begin{gather}\label{eq:storedenergyvariation_NandM}
        \bs{N}^\alpha = N^{\beta \alpha}\bs{a}_\beta + N^\alpha \bs{n}, \quad \bs{M}^{\alpha \beta} = M^{\lambda \alpha \beta}\bs{a}_{\lambda} + M^{\alpha\beta}\bs{n}, \\
\bs u_{|\alpha \beta} := \bs u_{,\alpha \beta} - \bar{\Gamma}^\gamma_{\alpha \beta} \bs u_{,\gamma},
\end{gather}
with 
\begin{gather}
    M^{\lambda \alpha \beta} = \frac{1}{2}\left( \frac{\partial \scrW}{\partial S_{\alpha \beta}^\mu} + \frac{\partial \scrW}{\partial S_{\beta \alpha}^\mu}\right)a^{\mu \lambda}, \qquad M^{\alpha \beta} = \frac{1}{2}\left(\frac{\partial \scrW}{\partial b_{\alpha \beta}} + \frac{\partial \scrW}{\partial b_{\beta \alpha }}\right). \nonumber \\
    N^{\beta \alpha} = \sigma^{\beta \alpha} - M^{\beta \gamma \mu}S^{\alpha}_{\gamma \mu}, \qquad N^{\alpha} = M^{\alpha \mu \gamma}b_{\mu\gamma} - M^{\beta \lambda} S_{\beta \lambda}^\alpha, \\
     \sigma^{\beta \alpha} = \frac{\partial \scrW}{\partial a_{\alpha\beta}} + \frac{\partial \scrW}{\partial a_{\beta \alpha}}. \label{eq:storedenergyvariation_NandM_decomposition}
\end{gather}

Using the above, we may re-express $\dot{E}$ as
\begin{align}
    \label{eq:energyvariation}
    \dot{E} = \int_{\partial \Omega} \varphi^\alpha \nu_\alpha \, ds - \int_\Omega \bs{u} \cdot \bs{T}_{|\alpha}^\alpha \, dA,
\end{align}
where
\begin{gather}
    \varphi^\alpha = \bs{T}^\alpha \cdot \bs{u} + \bs{M}^{\alpha \beta} \cdot \bs{u}_{,\beta}, \quad 
    \bs{T}^\alpha = \bs{N}^\alpha - \bs{M}^{\alpha \beta}_{|\beta}, \quad \bs T^\alpha_{|\alpha} := \frac{1}{\sqrt{A}}(\sqrt{A} \bs T^\alpha)_{,\alpha}, \\
    \bs M^{\alpha \beta}_{|\gamma} := \bs M^{\alpha \beta}_{,\gamma} + \bs M^{\alpha \lambda} \bar{\Gamma}^\beta_{\lambda \gamma} + \bs M^{\lambda \beta} \bar{\Gamma}^\alpha_{\lambda \gamma},
\end{gather}
and
\begin{align}
    \nu_\alpha = \bs{\nu} \cdot \bs{A}_\alpha
\end{align}
are the covariant components of the outward unit normal vector to the boundary, $\bs{\nu}$. From here, it follows from integration by parts techniques \cite{Steigmann18Lattice} that the boundary integral may be recast as
\begin{align}
    \label{eq:boundaryibp}
    \int_{\partial \Omega} \varphi^\alpha \nu_\alpha \, ds & = \int_{\partial \Omega} \left[ \left(\bs{T}^\alpha \nu_\alpha - (\bs{M}^{\alpha \beta} \nu_\alpha \tau_\beta)' \right) \cdot \bs{u} + \bs{M}^{\alpha \beta} \nu_\alpha \nu_\beta \cdot \bs{u}_{\nu} \right] \, ds \\ & \quad - \sum [\bs{M}^{\alpha \beta} \nu_\alpha \tau_\beta]_i \cdot \bs{u}_i,
\end{align}
where $\tau_\alpha = \bs{\tau} \cdot \bs{A}_{\alpha}$ are the covariant components for the unit tangent to the boundary, $\bs{\tau}$, $\bs{u}_\nu$ is the normal derivative (with respect to $\bs \nu$), $(\cdot)'$ denotes differentiation with respect to the arc-length parameter $s$ on $\partial \Omega$, and the brackets with subscript $[\cdot ]_i$ represent the jump in the argument along the $i$th corner of the boundary.

The virtual work expended on the shell is assumed to be of the form
\begin{align}
    P = \int_\Omega \bs{g} \cdot \bs{u} \, dA + \int_{\partial \Omega_t} \bs{t} \cdot \bs{u} \, ds + \int_{\partial \Omega_m} \bs{m} \cdot \bs{u}_\nu \, ds + \sum_{i=1}^N \bs{f}_i \cdot \bs{u}_i,
\end{align}
where $\bs{g}$ is the body force density, $\bs{t}$ is the edge traction assigned on $\partial \Omega_t \subset \partial \Omega$, $\bs{m}$ is the edge double force assigned on $\partial \Omega_m  \subset \p \Omega$, and $\bs{f}_i$ is the corner force assigned at the $i$th corner (we assume that both force/double force/corner force and position are not assigned at the same points throughout). Taken with \eqref{eq:energyvariation} and \eqref{eq:boundaryibp}, the form of $P$ and the fundamental lemma of the calculus of variations imply that the equilibrium equations for the shell are
\begin{align}
    \bs{T}^\alpha_{|\alpha} + \bs g = \bs 0.
\end{align}    
Moreover, the edge traction, edge double force and corner force satisfy
\begin{gather}
    \quad \bs{t} = \bs{T}^\alpha \nu_\alpha - (\bs{M}^{\alpha \beta} \nu_\alpha \tau_\beta)', \quad 
    \bs{m} = \bs{M}^{\alpha \beta} \nu_a \nu_b, \quad \bs{f}_i = - [\bs{M}^{\alpha \beta} \nu_\alpha \tau_\beta ]_i.  
\end{gather}
We recall from Section 5.3 of \cite{Steig13} that
\begin{align}
    \bs c := \bs y_{,\nu} \times \bs m
\end{align}
can be interpreted as a density of edge couples.

Going forward, we now assume for simplicity that the shell energy $\scrW$ is given by the leading cubic-in-order expression appearing in \eqref{eq:dilatsurfaceenergy},
\begin{align*}
\scrW 
&= W_{\mathrm{Koiter}} + {h\ell_s^2 \mu}\frac{2 \mu^2}{(\lambda + 2\mu)^2}\bigl ( |\nabla_s (\tr \bs \eps)|^2 + |\tr \bs \rho|^2 \bigr ) \\
&= h \Bigl (
	\frac{\lambda \mu}{\lambda + 2\mu}(\tr \bs \eps)^2 + \mu |\bs \eps|^2
	\Bigr ) + \frac{h^3}{24} \Bigl ( \frac{\lambda \mu}{\lambda + 2\mu}(\tr \bs \rho)^2 + \mu |\bs \rho|^2 \Bigr )\\ 
&\quad + {h\ell_s^2 \mu}\frac{2 \mu^2}{(\lambda + 2\mu)^2}\bigl ( |\nabla_s (\tr \bs \eps)|^2 + |\tr \bs \rho|^2 \bigr ).\label{eq:dilsurfen}
\end{align*}
We now record some identities that will be useful for determining $\bs T^\al$, $\bs t$, $\bs m$, and $\bs f_i$. In what follows, the functions differentiated with respect to $a_{\sigma \tau}, b_{\sigma \tau},$ and $S^\lambda_{\sigma \tau}$ are written in terms of these variables in a way that is symmetric under interchanging $\sigma$ and $\tau$, and the components of various tensors are raised using the reference dual metric components.  

For $\bs{\eps}$, we have 
\begin{align*}
    \frac{\partial \eps_{\alpha \beta}}{\partial a_{\sigma \tau}} &= \frac{1}{4} \left( \delta_\alpha^\sigma \delta_\beta^\tau + \delta_\alpha^\tau \delta_\beta^\sigma \right),  \quad \frac{\partial \eps_{\alpha \beta}}{\partial b_{\sigma \tau}} = \frac{\partial \eps_{\alpha \beta}}{\partial S_{\sigma \tau}^\lambda} = 0,\\
    \frac{\partial (\tr \bs{\eps})^2}{\partial a_{\sigma \tau}} &= A^{\sigma \tau} \tr{\bs{\eps}}, \quad \frac{\partial (\tr \bs{\eps})^2}{\partial b_{\sigma \tau}} = \frac{\partial (\tr \bs{\eps})^2}{\partial S^{\lambda}_{\sigma \tau}} =0,
    \\
    \frac{\partial | \bs{\eps}|^2}{\partial a_{\sigma \tau}} &= \eps^{\sigma \tau}, \quad \frac{\partial | \bs{\eps}|^2}{\partial b_{\sigma \tau}} = \frac{\partial | \bs{\eps}|^2}{\partial S^{\lambda}_{\sigma \tau}} =0.
\end{align*}
Similarly, for $\bs{\rho}$, we have 
\begin{align*}
    \frac{\partial \rho_{\alpha \beta}}{\partial b_{\sigma \tau}} &= \frac{1}{2} \left( \delta_\alpha^\sigma \delta_\beta^\tau + \delta_\alpha^\tau \delta_\beta^\sigma \right),  \quad \frac{\partial \rho_{\alpha \beta}}{\partial a_{\sigma \tau}} = \frac{\partial \rho_{\alpha \beta}}{\partial S_{\sigma \tau}^\lambda} = 0,\\
    \frac{\partial (\tr \bs{\rho})^2}{\partial b_{\sigma \tau}} &= 2A^{\sigma \tau} \tr{\bs{\rho}}, \quad \frac{\partial (\tr \bs{\rho})^2}{\partial a_{\sigma \tau}} = \frac{\partial (\tr \bs{\rho})^2}{\partial S^{\lambda}_{\sigma \tau}} =0,
    \\
    \frac{\partial | \bs{\rho}|^2}{\partial b_{\sigma \tau}} &= 2\rho^{\sigma \tau}, \quad \frac{\partial | \bs{\rho}|^2}{\partial a_{\sigma \tau}} = \frac{\partial | \bs{\rho}|^2}{\partial S^{\lambda}_{\sigma \tau}} =0.
\end{align*}
To compute the partial derivatives of $a_{\alpha \beta, \gamma}$, observe that 
\begin{align*}
        a_{\alpha \beta , \gamma} & = \bs{a}_{\alpha, \gamma} \cdot \bs{a}_{\beta} + \bs{a}_\alpha \cdot \bs{a}_{\beta, \gamma} \\
        & = \Gamma_{\alpha \gamma}^\delta \bs{a}_\delta \cdot \bs{a}_\beta + \bs{a}_\alpha \cdot \Gamma_{\beta \gamma}^\delta \bs{a}_\delta \\
        & = \Gamma_{\alpha \gamma}^\delta a_{\delta \beta} + \Gamma_{\beta \gamma}^\delta a_{\alpha \delta} \\
        & = \left( S_{\alpha \gamma}^\delta + \overline{\Gamma}_{\alpha \gamma}^\delta\right) a_{\delta \beta} + \left( S_{\beta \gamma}^\delta + \overline{\Gamma}_{\beta \gamma}^\delta\right) a_{\alpha \delta}.
    \end{align*}
    This yields 
    \begin{align*}
        \frac{\partial a_{\alpha \beta, \gamma}}{\partial a_{\sigma \tau}} &= \frac{1}{2} \left( \Gamma_{\alpha \gamma}^\sigma \delta_\beta^\tau + \Gamma_{\alpha \gamma}^\tau \delta_\beta^\sigma \right) + \frac{1}{2} \left( \Gamma_{\beta \gamma}^\tau \delta_\alpha^\sigma + \Gamma_{\beta \gamma}^\sigma \delta_\alpha^\tau \right), \\
        \frac{\partial a_{\alpha \beta, \gamma}}{\partial S^{\lambda}_{\sigma \tau}} &= \frac{1}{2} \left( \delta_\alpha^\sigma \delta_\gamma^\tau + \delta_\alpha^\tau \delta_\gamma^\sigma \right) a_{\lambda \beta} + \frac{1}{2} \left( \delta_\beta^\sigma \delta_\gamma^\tau + \delta_\beta^\tau \delta_\gamma^\sigma \right) a_{\alpha \lambda}, \\
        \frac{\partial a_{\alpha \beta, \gamma}}{\partial b_{\sigma \tau}} & = 0.
    \end{align*}
    Lastly, to compute the partial derivatives of $|\nabla_s \tr \bs \eps|^2$, we have by definition $
        \nabla_s \tr \bs\eps = (\tr \bs\eps)_{,\gamma} \bs{A}^\gamma$, so 
        \begin{align*}
            \frac{\partial |\nabla_s \tr \bs \eps|^2}{\partial a_{\sigma \tau}} & = 2 (\tr \bs \eps)_{, \delta} A^{\gamma \delta}\frac{\partial (\tr \bs \eps)_{,\gamma}}{\partial a_{\sigma \tau}}.
        \end{align*}
    Since $(\tr \bs \eps)_{,\gamma} = \eps_{\alpha\beta,\gamma}A^{\alpha\beta} + \eps_{\alpha\beta}A^{\alpha \beta}_{,\gamma}$, we can use previous computations to find that
    \begin{align}
        \frac{\partial(\tr \bs \eps)_{,\gamma}}{ \partial a_{\sigma \tau }} & = \frac{1}{2} \frac{\partial a_{\alpha \beta, \gamma}}{\partial a_{\sigma \tau}} A^{\alpha \beta} + \frac{\partial \eps_{\alpha \beta}}{\partial a_{\sigma \tau}}  A^{\alpha \beta}_{,\gamma} \\
        & = \frac{1}{2} \left(S^{\sigma}_{\alpha \gamma} + \overline{\Gamma}^\sigma_{\alpha \gamma} \right) A^{\alpha \tau} + \frac{1}{2} \left( S^{\tau}_{\alpha \gamma} + \overline{\Gamma}^\tau_{\alpha \gamma} \right) A^{\alpha \sigma} + \frac{1}{2} A^{\sigma \tau}_{,\gamma}.
    \end{align}
    Hence,
    \begin{align}
            \frac{\partial |\nabla_s \tr \bs \eps|^2}{\partial a_{\sigma \tau}} & =  (\tr \bs \eps)_{, \delta} A^{\gamma \delta} \left[ 
            \left(S^{\sigma}_{\alpha \gamma} + \overline{\Gamma}^\sigma_{\alpha \gamma} \right) A^{\alpha \tau} + \left( S^{\tau}_{\alpha \gamma} + \overline{\Gamma}^\tau_{\alpha \gamma} \right) A^{\alpha \sigma} + A^{\sigma \tau}_{,\gamma} \right].
        \end{align}
        Similarly
        \begin{align}
            \frac{\partial |\nabla_s \tr \bs \eps|^2}{ \partial S^{\lambda}_{\sigma \tau}} = (\tr \bs{\eps})_{,\delta} \left[ A^{\tau \delta} A^{\sigma \alpha} + A^{\tau \alpha} A^{\sigma \delta} \right] a_{\alpha \lambda}, \quad \frac{\partial |\nabla_S \tr \bs \eps|^2}{\partial b_{\sigma \tau}} =0. 
        \end{align}

We incorporate all of the above into explicit computations of $\sigma^{\sigma \tau}$, $M^{\sigma \tau}$, and $M^{\mu \sigma \tau}$ for $\scrW$ given by \eqref{eq:dilsurfen}: 
\begin{align}
    \sigma^{\sigma \tau}  
    & = h \left[ \frac{{2}\lambda \mu}{\lambda + 2 \mu} A^{\sigma \tau} \tr \bs{\eps} + {2}\mu \eps^{\sigma \tau} \right] \\ \quad 
    &\quad+ h l_s^2 \mu \frac{{4} \mu^2}{(\lambda + 2 \mu)^2} A^{\gamma \delta} (\tr \bs{\eps})_{,\delta} \left[  \Gamma^\sigma_{\alpha \gamma}  A^{\alpha \tau} +  \Gamma^\tau_{\alpha \gamma} A^{\alpha \sigma}  + A^{\sigma \tau}_{,\gamma} \right], \label{eq:sigma}\\
    M^{\sigma \tau} 
    & = \frac{h^3}{12} \left[ \frac{\lambda \mu}{\lambda + 2 \mu} A^{\sigma \tau} \tr \bs{\rho} + \mu \rho^{\sigma \tau} \right] + h l_s^2 \mu \frac{4\mu^2}{(\lambda + 2\mu)^2} A^{\sigma \tau} \tr \bs{\rho}, \label{eq:M} \\
    M^{\lambda \sigma \tau} 
    & = h l_s^2 \mu \frac{2 \mu^2}{(\lambda + 2\mu)^2} (\tr \bs{\eps})_{,\delta} \left( A^{\tau \delta} A^{\sigma \lambda} + A^{\sigma \delta} A^{\tau \lambda} \right). \label{eq:MM}
\end{align}
The above formulae will now be used to determine the body force $\bs g$, edge tractions $\bs t$, and edge double force densities $\bs m$ necessary to support equilibrium states including a finite plate deformed into a cylinder, combined extension/compression and radial expansion/contraction of a finite cylinder, and pure torsion of a finite cylinder. Since the reference metric is flat in all settings, covariant, contravariant and mixed components of various tensors coincide. 

\subsection{Rolling a finite flat plate into a cylinder} 
We assume that the shell is initially a finite flat plate, $\Omega = [-R\pi, R\pi] \times [0,L] \times \{0\}$. We choose $(\theta^1, \theta^2):= (R\theta, z)$ with $(\theta, z) \in [-\pi, \pi] \times [0, L]$, normal vector $\bs e_3$, and 
$$\bs{x} = \theta^\alpha \bs{e}_{\alpha}.$$
We consider the deformation of isometrically rolling the plate into a right circular cylinder, 
$$\bs{y} = R\bs{e}_{r}(\theta) + z\bs{k},$$ where $\bs{e}_r(\theta) = \cos\theta \bs{e}_1 + \sin\theta\bs e_{2}$, $\bs{k} = \bs{e}_3$ and $\bs{e}_\theta = \bs{k} \times \bs{e}_r$. We observe that $\bs{A}_\alpha = \bs{e}_\alpha, \bs{N} = \bs{e}_3$ and $\bs{a}_1 = \bs{e}_\theta, \bs{a}_2 = \bs{k}, \bs{n} = \bs{a}_1 \times \bs{a}_2 = \bs{e}_r.$ Furthermore, for all $\al, \beta, \mu$,
\begin{gather}
\eps_{\al \beta} = 0, \quad b_{\al \beta} = -R^{-1}\delta_{\al 1}\delta_{\beta 1}, \quad S^\mu_{\alpha \beta} = 0. 
\end{gather}
We see from \eqref{eq:sigma}, \eqref{eq:M}, and \eqref{eq:MM} that for all $\alpha, \beta$
\begin{gather}
N^{\beta\alpha} = \sigma^{\beta\alpha} = 0, \quad N^\alpha = 0, \implies \bs{N}^{\alpha} =\bs{0}, 
\end{gather}
and moreover, $\bs{M}^{\alpha\beta} = M^{\alpha \beta}\bs{e}_{r}(\theta)$ with $\bs M^{12} = \bs M^{21} = \bs 0$ and 
\begin{align}
     \bs{M}^{11} &= -R^{-1}\left(\frac{h^3}{12} \left[ \frac{\lambda \mu}{\lambda + 2\mu} + \mu \right] + h l_s^2 \mu \frac{4\mu^2}{(\lambda + 2\mu)^2}\right)\bs{e}_r, \\
     \bs{M}^{22} &= -R^{-1}\left(\frac{h^3}{12} \left[ \frac{\lambda \mu}{\lambda + 2\mu} \right] + h l_s^2 \mu \frac{4\mu^2}{(\lambda + 2\mu)^2}\right)\bs{e}_r.
\end{align}
In particular, 
\begin{align}
\bs T^\alpha = - \bs M^{\alpha \beta}_{,\beta}. 
\end{align}
For the body force, we compute
\begin{align}
    \bs{g} = -\bs{T}^{\alpha}_{,\alpha} = \bs{M}^{\alpha \beta}_{,\alpha \beta} = \bs M^{11}_{,11} = -R^{-2}  M^{11} ,
\end{align}
and thus, an outward radial force must be applied at every point of the cylinder for equilibrium to hold. Next, we compute the traction on the boundary components. For $\{z =0 \}$, we have that $\bs{\tau} = \tau_{\alpha}\bs{A}^\alpha = \tau_{\alpha}\bs{e}_{\alpha} = \bs{e}_{1}$ and $\bs{\nu} = \nu_{\alpha}\bs{e}_{\alpha} = -\bs{e}_2$. Hence, we see that 
\begin{align}
    \bs{t} &= \bs T^\al \nu_\al = -\bs T^2 = \bs M^{2\beta}_{,\beta} = 0, \\
    \bs{m} &= \bs{M}^{22}\\
   \bs{c}& = {\bs y_{,\nu} \times \bs m = -\bs k \times  M^{22}\bs e_r = -M^{22} \bs e_{\theta}}.
\end{align}
{ The vector field $\bs{c}$ is the density of edge couples preventing the lip of the cylinder from bowing out.}  
The same result is true for $\{z = L\}$, {with $\bs c= M^{22}\bs{e_\theta}$}.  For $\{\theta = \pm\pi\}$, we have $\bs{\tau} = \pm \bs{e}_2, \bs{\nu} = \pm \bs{e}_1$. Hence 
\begin{align}
     \bs{t} &= \mp \bs{M}^{11}_{,1} = \pm R^{-2}\left(\frac{h^3}{12} \left[ \frac{\lambda \mu}{\lambda + 2\mu} + \mu \right] + h l_s^2 \mu \frac{4\mu^2}{(\lambda + 2\mu)^2}\right)\bs{e}_\theta, \\
    \bs{m} &= \bs{M}^{11}\\
    \bs{c} &= { \pm \bs y_{,1} \times \bs m = \pm \bs e_{\theta} \times M^{11} \bs e_r = \mp M^{11}\bs k }.
\end{align} 
{ Thus, there is an edge force pulling the two edges together, along with a bending couple density $\bs c$ parallel to the axis $\bs k$.} The corner forces are all $0$. 
\subsection{Combined extension/compression and radial expansion/contraction}
We now assume that $\bs{x} = R\bs{e}_r(\theta) + z\bs{k}$ and $\bs{y} = \eta R \bs{e}_{r}(\theta) + \zeta z \bs{k}$ with $\eta, \zeta >0$. It follows from above that $\bs{A}_{1} = \bs{e}_\theta, \bs{A}_2 = \bs{k}, \bs{N} = \bs{e}_r,$ and $\bs{a}_1 = \eta \bs{e}_\theta, \bs{a}_2 = \zeta \bs{k}, \bs{n} = \bs{e}_r.$ We compute 
\begin{align}
    (a_{\alpha \beta}) = \begin{pmatrix}
        \eta^2  & 0 \\
        0 & \zeta^2
    \end{pmatrix}, \quad  
    (b_{\alpha \beta})= \begin{pmatrix}
        \frac{-\eta}{R} & 0 \\
        0 & 0
    \end{pmatrix}, \quad \Gamma^{\mu}_{\alpha \beta} = \bar \Gamma^\mu_{\alpha \beta} =0,
\end{align}
which implies that 
\begin{align}
    (\eps_{\alpha \beta}) = \begin{pmatrix}
        \frac{\eta^2 -1}{2} & 0 \\
        0 & \frac{\zeta^2 -1}{2}
    \end{pmatrix}, \quad 
    (\rho_{\alpha \beta}) = \begin{pmatrix}
        \frac{1 - \eta}{R} & 0 \\
        0 & 0
    \end{pmatrix}.
\end{align}

We compute that
\begin{align}
    N^{11} & = 2 h \mu \frac{\lambda + \mu}{\lambda + 2\mu} \left( \eta^2 -1 \right) + h \frac{\lambda \mu}{\lambda + 2 \mu} \left( \zeta^2 - 1 \right) ,\\
    N^{22} & = h \frac{\lambda \mu}{\lambda + 2 \mu} \left( \eta^2 - 1 \right) + 2 h \mu \frac{\lambda + \mu}{\lambda + 2\mu} \left( \zeta^2 -1 \right),\\
    M^{11} & = \left( \frac{h^3\mu }{6} \frac{\lambda + \mu}{\lambda + 2\mu} + h l_s^2 \mu \frac{4\mu^2}{(\lambda + 2\mu)^2}\right) \frac{1-\eta}{R},\\
    M^{22} & = \left( \frac{h^3}{12} \frac{\lambda \mu}{\lambda + 2\mu} + h l_s^2 \mu \frac{4\mu^2}{(\lambda + 2\mu)^2}\right) \frac{1-\eta}{R},
\end{align}
with the off-diagonal components of $\sigma$ and $M$ vanishing. Additionally $M^{\lambda \alpha \beta} = 0$ for all $\lambda, \alpha, \beta$. From here, we find
\begin{align}
    \bs{T}^1 = \left( \eta N^{11} - \frac{M^{11}}{ R} \right) \bs{e}_\theta, \quad \bs{T}^2 = \zeta N^{22}\bs{k},
\end{align}
which immediately implies that the required body force is
\begin{gather*}
    \bs{g} = -\bs T^{\alpha}_{,\alpha} = \left( \frac{\eta N^{11}}{R} - \frac{M^{11}}{R^2}\right) \bs{e}_r.
\end{gather*}

When $z=0$, $\bs{\nu} = -\bs{k}$ and $\bs{\tau} = \bs{e}_\theta$, so $\bs{t} |_{z=0} = -\bs{T}^2 = -N^{22} \zeta \bs{k}$. On the other edge, where $z = L$, $\bs{\nu} = \bs{k}$, so $\bs{t} |_{z=L} = \bs{T}^2 = N^{22} \zeta \bs{k}$. On both edges, $\bs{m} = M^{22} \bs{e}_r${, which implies that $\bs{c}|_{z = 0} = -\zeta \bs{k} \times  M^{22}\bs e_r = - \zeta M^{22} \bs{e_\theta}$ and $\bs{c}|_{z=L} = \zeta M^{22}\bs e_{\theta}$.}

In the case of materials with $\lambda, \mu > 0$ (i.e., Poisson's ratio is positive), we can consider the semi-inverse problem of supposing $\eta < 1$ and $\zeta > 1$ with $\bs{g} = \bs 0$. This corresponds to increasing the height of the cylinder and shrinking its radius while applying no body forces. For any $\eta < 1$, a unique $\zeta > 1$ satisfying this requirement always exists. On the $z=L$ edge of the cylinder, we may interpret the traction as a sum of two forces: one depending on $\zeta$ in the $+ \bs{k}$ direction that pulls the edges apart to increase the height and one depending on $\eta$ that applies a slight compression in the $-\bs{k}$ direction to counteract the natural stretching of the cylinder as the radius decreases. {Furthermore, we may interpret $\bs{c}$ as a density of edge couples that prevents the edges of the cylinder from bowing inwards during the deformation.}

\subsection{Pure torsion}
For $\gamma \in \bbR$, set $\vartheta := \theta + \frac{\gamma}{R}z$. We now consider the deformation of $\bs{x} = R\bs{e}_r(\theta) + z\bs{k}$ to $\bs{y} = R\bs{e}_r(\vartheta) + z \bs{k}$. It follows that $\bs{a}_1 = \bs{e}_\theta(\vartheta), \bs{a}_2 = \gamma \bs{e}_\theta(\vartheta) + \bs{k}, \bs{n} = \bs{e}_r(\vartheta).$ Then 
\begin{align}
    (a_{\al \beta}) = \begin{pmatrix}
        1 & \gamma \\
        \gamma & \gamma^2 + 1
    \end{pmatrix}, \quad (b_{\al \beta}) = -\frac{1}{R}\begin{pmatrix}
        1 & \gamma \\
        \gamma & \gamma^2
    \end{pmatrix}, \quad \Gamma^\mu_{\alpha\beta} = \bar{\Gamma}^\mu_{\alpha \beta}=0,
\end{align}
which implies that 
\begin{align}
    (\eps_{\al \beta}) = \frac{1}{2}\begin{pmatrix}
        0 & \gamma\\
        \gamma & \gamma^2
    \end{pmatrix}, \quad 
    (\rho_{\al \beta}) = -\frac{1}{R}\begin{pmatrix}
        0 & \gamma \\
        \gamma & \gamma^2
    \end{pmatrix}.
\end{align}
We can now compute $\bs{N}^{\alpha} = N^{\alpha\beta}\bs{a}_{\beta}$ with
\begin{align*}
    N^{11} &= h\gamma^2 \frac{\lambda \mu}{\lambda + 2\mu},\\
    N^{12} &=N^{21}= h\gamma \mu,\\
    N^{22} & =  
    h\gamma^2\left[\frac{\lambda \mu}{\lambda + 2\mu}  + \mu \right],
\end{align*}
and $\bs{M}^{\alpha\beta} = M^{\alpha\beta}\bs e_r(\vartheta)$ with 
\begin{align}
    M^{11} &= -\frac{\gamma^2}{R}\left(\frac{h^3}{12}\frac{\lambda \mu}{\lambda + 2\mu} + hl_s^2 \mu \frac{4\mu^2}{(\lambda + 2\mu)^2}\right),\\
    M^{12}&=M^{21} = -\frac{\gamma}{R}\frac{h^3}{12}\mu,\\
    M^{22} &= -\frac{\gamma^2}{R}\left(\frac{h^3}{12}\left[\frac{\lambda \mu}{\lambda + 2\mu} + \mu\right]  + hl_s^2\mu \frac{4\mu^2}{(\lambda + 2\mu)^2}\right).
\end{align}
From here we see that 
\begin{align}
    \bs T^1  &= \left( N^{11} + \gamma N^{21} - \frac{1}{R}(M^{11} + \gamma M^{12})\right) \bs e_{\theta}(\vartheta) + N^{21}\bs k,\\
    \bs T^2 
    &= \left(N^{12} + \gamma N^{22} - \frac{1}{R}(M^{21} + \gamma M^{22}\right) \bs e_{\theta}(\vartheta)+ N^{22}\bs k.
\end{align}
It follows from this that 
\begin{align}
    \bs{g}
    & = \left[-\frac{1}{R^2}\left(M^{11}+ 2\gamma M^{12} + \gamma^2 M^{22} \right)  + \frac{1}{R} \left(N^{11} + 2\gamma N^{12} + \gamma^2 N^{22}\right)\right] \bs{e}_r(\vartheta).
\end{align}
On the boundaries $\{z = 0\}$ and $\{z = L\}$, as in the previous example we have that $\bs t |_{z = 0} = -\bs T^2, $ $\bs t |_{z = L}  = \bs T^2$ and on both edges, $\bs m = M^{22} \bs{e}_r(\vartheta)$.{ This implies that 
\begin{align}
    \bs{c}|_{z = 0} &= -\bs a_2 \times \bs m 
    = -(\gamma \bs e_{\theta} + \bs k ) \times M^{22} \bs{e}_r = \gamma M^{22}\bs k + -M^{22}\bs{e}_{\theta}\\
    \bs{c}|_{z = L} & = -\gamma M^{22}\bs{k} + M^{22}\bs{e}_\theta.
\end{align} }Thus, when $\lambda, \mu>0$ (i.e., Poisson's ratio is positive), we must apply an outward radial force to the cylinder which prevents the cylinder from crushing inward during torsion. We see also that we must twist in the  direction of the torsion and pull outward on the cylinder to prevent the cylinder from twisting back or crushing along $\bs k$ under the twist. Finally, the value of $\bs{c}$ shows that density of edge couples has components in both the $\bs k$ and $\bs e_\theta$ directions that prevents the cylinder lip from bowing out in a skew direction.

\bibliographystyle{plain}
\bibliography{researchbibmech}

\bigskip
\footnotesize

\noindent \textsc{Department of Mathematics, University of North Carolina at Chapel Hill, CB 3250 Phillips Hall,
Chapel Hill, NC 27599}\\
\noindent \textit{E-mail address}: \texttt{corbindb@unc.edu}
\bigskip

\noindent \textsc{Department of Mathematics, University of North Carolina at Chapel Hill, CB 3250 Phillips Hall,
Chapel Hill, NC 27599}\\
\noindent \textit{E-mail address}: \texttt{canzani@email.unc.edu}

\bigskip 

\noindent \textsc{Department of Mathematics, University of North Carolina at Chapel Hill, CB 3250 Phillips Hall,
Chapel Hill, NC 27599}\\
\noindent \textit{E-mail address}: \texttt{scott\_hallyburton@unc.edu}

\bigskip

\noindent \textsc{Department of Mathematics, University of North Carolina at Chapel Hill, CB 3250 Phillips Hall,
Chapel Hill, NC 27599}\\
\noindent \textit{E-mail address}: \texttt{jrmott@email.unc.edu}

\bigskip

\noindent \textsc{Department of Mathematics, University of North Carolina at Chapel Hill, CB 3250 Phillips Hall,
Chapel Hill, NC 27599}\\
\noindent \textit{E-mail address}: \texttt{crodrig@email.unc.edu}

\end{document}